\documentclass[aps,twocolumn,showpacs,amsmath,amssymb,superscriptaddress,floatfix]{revtex4-1}

\usepackage{graphicx}
\usepackage{dcolumn}
\usepackage{bm}
\usepackage{amssymb,mathtools}
\usepackage{booktabs}
\usepackage{color}
\usepackage{microtype}
\usepackage[T1]{fontenc}
\usepackage{lmodern} 
\usepackage{enumitem}

\usepackage{physics}
\usepackage{upgreek}

\allowdisplaybreaks
\usepackage{placeins}

\usepackage[colorlinks,plainpages=false,linkcolor=blue,urlcolor=blue,citecolor=blue,pdfpagemode=UseNone,pdfstartview=FitBH]{hyperref}

\usepackage{sidecap}

\renewcommand{\tablename}{Table}
\makeatletter\renewcommand{\fnum@table}[1]{\tablename~\thetable.}\makeatother

\makeatletter

\def\titlecase#1{%
\let\tc@w\@empty
\protected@edef\tmp{\noexpand\tc@a#1\relax}\expandafter\tc@uc@\tmp}

\def\tc@a{\futurelet\tmp\tc@aa}

\def\tc@aa{%
\ifcat a\noexpand\tmp\expandafter\tc@ab
\else\expandafter\tc@ac\fi}

\def\tc@ab#1{\edef\tc@w{\tc@w#1}\tc@a}

\def\tc@ac{%
\csname tc@@\tc@w\endcsname\expandafter\tc@uc\tc@w
\let\tc@w\@empty
\ifx\tmp\@sptoken\let\next\tc@sp
\else\ifx\tmp\relax\let\next\relax
\else\let\next\tc@nxt
\fi\fi\next}

\def\tc@sp#1{ \tc@a#1}
\def\tc@nxt#1{#1\tc@a}

\def\tc@uc#1{\uppercase{#1}}
\def\tc@uc@#1#2{\uppercase{#1#2}}

\let\tc@@the\@gobbletwo
\let\tc@@and\@gobbletwo

\makeatother

\newcommand{\ii}{\mathrm{i}}

\renewcommand{\Im}{\operatorname{Im}}
\newcommand{\overbar}[1]{\mkern 1.5mu\overline{\mkern-1.5mu#1\mkern-1.5mu}\mkern 1.5mu}

\DeclareSymbolFont{bbold}{U}{bbold}{m}{n}
\DeclareSymbolFontAlphabet{\mathbbm}{bbold}

\newcommand{\h}[1]{H_\mathrm{#1}}

\begin{document}

\title{Natural-Orbital Impurity Solver and Projection Approach for Green's Function}

\author{Y.~Lu}
\email{y.lu@thphys.uni-heidelberg.de}
\affiliation{Institute for Theoretical Physics, Heidelberg University, Philosophenweg 19, 69120 Heidelberg, Germany}

\author{X.~Cao}
\affiliation{Max Planck Institute for Solid State Research, Heisenbergstrasse~1, 70569 Stuttgart, Germany}

\author{P.~Hansmann}
\affiliation{Max Planck Institute for Solid State Research, Heisenbergstrasse~1, 70569 Stuttgart, Germany}
\affiliation{Max Planck Institute for Chemical Physics of Solids, N\"othnitzer Strasse 40, 01187 Dresden, Germany\looseness=-1}

\author{M.~W.~Haverkort}
\affiliation{Institute for Theoretical Physics, Heidelberg University, Philosophenweg 19, 69120 Heidelberg, Germany}

\date{\today}
\pacs{}

\begin{abstract}
  We extend a previously proposed rotation and truncation scheme to optimize quantum Anderson impurity calculations with exact diagonalization [PRB \textbf{90}, 085102 (2014)] to density-matrix renormalization group (DMRG) calculations. The method reduces the solution of a full impurity problem with virtually unlimited bath sites to that of a small subsystem based on a natural impurity orbital basis set. The later is solved by DMRG in combination with a restricted-active-space truncation scheme. The method allows one to compute Green's functions directly on the real frequency or time axis. We critically test the convergence of the truncation scheme using a one-band Hubbard model solved in the dynamical mean-field theory. The projection is exact in the limit of both infinitely large and small Coulomb interactions. For all parameter ranges the accuracy of the projected solution converges exponentially to the exact solution with increasing subsystem size.
\end{abstract}

\maketitle

\section{Introduction}

The class of quantum impurity models are of long-standing interest to physicists. They describe a wide range of quantum mechanical problems that involve a subsystem with a limited number of degrees of freedom (an impurity) coupled to a much larger system (a bath) that contains a quasi-continuum of degrees of freedom. Examples include the Kondo and heavy-fermion systems~\cite{Gunnarsson:1983uf,Stewart1984,Hewson1997}, core-level X-ray spectroscopy~\cite{vdL1986,Haverkort2014}, tunneling in dissipative systems~\cite{Leggett1987}, and various problems in quantum optics~\cite{John1990}. In the past years, the interest in impurity models has also been reinvigorated by the continuous development of dynamical mean-field theory (DMFT)~\cite{Metzner1989,Georges1996}, in which the correlated lattice problem is mapped self-consistently to an effective impurity model. DMFT allows for exact treatment of the local electronic correlations and has proven to correctly describe the electronic structure of many strongly correlated materials, which was beyond the reach of traditional mean-field or independent-particle methods.

At the core of DMFT, or an impurity model in general, is the efficient and accurate solution of the impurity ground state and one-body Green's functions. To this end, many numerical methods have been developed, including the quantum Monte Carlo (QMC)~\cite{Gull2011,Georges1992,Ulmke1995,Rubtsov2005,Werner2006,Werner2006b}, numerical renormalization group (NRG)~\cite{Wilson1975,Bulla2008,Bulla1999,Bulla2011,Bulla2005,Pruschke2000}, density-matrix renormalization group (DMRG)~\cite{White1992,White1993,Schollwock2005,Hallberg2006,Schollwock2011,Garcia2004,Raas2004,Wolf2014,Wolf2015,Ganahl2014,Ganahl2015,Holzner2010,Bauernfeind2017}, and exact diagonalization (ED)~\cite{Caffarel1994,Sangiovanni2006,Capone2007,Koch2008,Zgid2012,Lin2013,Lu2014}. Each method has its own merits and shortcomings. QMC can efficiently solve multi-band problems, yet by formulating on the imaginary axes, it entails an ill-conditioned inversion problem when obtaining real-frequency spectra~\cite{Jarrell1996,Lu2017}. In addition, its application to problems with low-symmetry interactions and/or off-diagonal Green's functions is often hindered by the fermionic sign problem. NRG is originally designed for impurity problems and works directly on the real axis. It has extremely good energy resolution for low-energy spectra. However, due to the necessarily logarithmic bath discretization~\cite{Bulla2005}, it lacks satisfactory resolution for high-energy features by construction. DMRG, when implemented on the real axis, treats all energies equally well, yet its solution is mostly limited to one and two band cases due to the exponential scaling of the complexity---or bond dimension in the matrix-product states (MPS) language---with the number of bands. It was shown recently that multi-band solution in DMRG is feasible by introducing fork tensor-product states~\cite{Holzner2010,Bauernfeind2017} as a variant of the conventional MPS. Another approach for countering the exponential growth of computation cost is to search for an optimized local basis for representing impurity problems. This has been most actively explored using ED methods~\cite{Zgid2012,Lu2014}, which are otherwise severely limited in accessible number of bands and bath sites. In Ref.~\cite{Lu2014}, some of us have demonstrated that a one-band impurity problem with a few hundred bath sites, ten times of that dealt in conventional ED, can be efficiently solved when represented on a natural-orbital basis set.

The optimized ED method above has also been tested in real, material-relevant scenarios involving general multi-orbital systems. It has been implemented in the freely available software package \textsc{Quanty}~\cite{Haverkort2012,Lu2014,Haverkort:2016hz}, which provides a flexible script language to solve quantum many-body problems. Several graphical interfaces are also available targeting specific spectroscopy calculations~\cite{Zimmermann:2018kw,retegan_crispy}, making efficient solutions to multi-orbital many-body impurity calculations accessible to a large audience (see references to \onlinecite{Haverkort2012,Lu2014,Haverkort:2016hz}). For generalized ligand-field theory calculations where an open $d$ or $f$ shell interacts with only a few ligand orbitals, the method works very well~\cite{Haverkort:2016hz,Agrestini:2017ft}. The same is true for observables that only need a limited resolution, e.g. several forms of core-level spectroscopy where the fine details of the spectra are smeared out by the large core-hole lifetime~\cite{Haverkort2014,Luder:2017fm}. For general materials, one often needs to correctly describe states with a bandwidth on the order of a Rydberg with a resolution better than the smallest energy scale (such as the crystal fields as small as tens of meV in some rare-earth compounds). Capturing all details of such materials requires one to have an energy resolution better than one per mille of the bandwidth. Such a requirement is crucial for understanding, for example, the detailed interaction of local orbital and crystal-field interactions with Kondo-like physics in some Ce compounds~\cite{Pourovskii:2014kk,Rueff:2015jr}. However, it is currently difficult to achieve for all impurity solver methods. Even for a single-band Hubbard model, capturing the exact line-shape of the onset of the Hubbard bands of a strongly correlated metal is still challenging~\cite{Karski:2005cm,Raas:2009jp,Granath:2012jl,Lu2014,Ganahl2014,Ganahl2015,Lee:2017kq}.

In this paper, we further explore the advantages of the natural-orbital representation of the impurity model, especially, by combining our method with DMRG. In practice, we exploit the energy separation of states provided by the natural-orbital representation, such that we can calculate the ground state and Green's functions of the impurity model (with a few hundred spin-orbitals) by projecting the full Hilbert space to a small subspace corresponding to low-order particle excitations. The projection scheme can be further simplified by constructing the projected states as product states of two subsystems, an interacting one containing the impurity site and a free one, respectively. Such a construction essentially allows for solution of the full impurity model by solving a small subsystem with only up to dozens of spin-orbitals. The proposed projection approach is applicable for all real-space wave-function based methods and can be straightforwardly implemented using ED and DMRG. In the following sections we show that the method, combined with the numerical advantages of DMRG, results in up to two orders of magnitude more efficient solution of a single-band Hubbard model in DMFT with improved accuracy when compared with our initial results obtained using ED in Ref.~\cite{Lu2014}.

\section{Natural-Orbital Impurity Solver}\label{sec:solver}

\begin{figure}[tb]
  \includegraphics{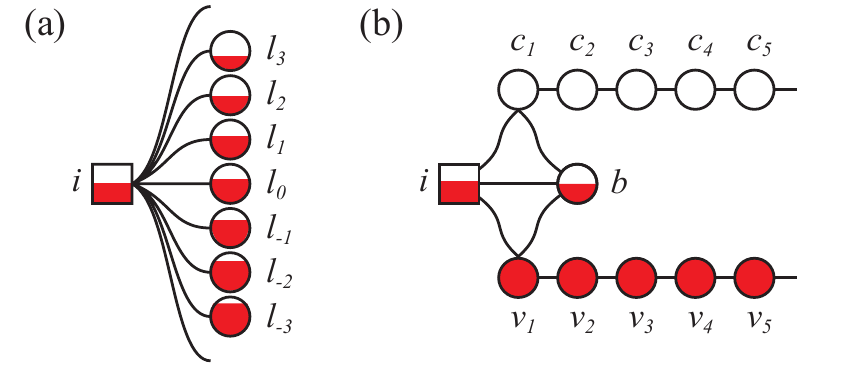}
  \caption{\label{fig:geo} Graphical representation of an impurity model in (a) the conventional ``star'' geometry and (b) the natural-orbital geometry (see text). The impurity is represented by a square and bath by circles. The solid lines represent hoppings between two sites. Each site consists of $m$ spin-orbitals.}
\end{figure}

A general Anderson impurity model is described by the Hamiltonian $\h{A}$ that contains two parts

\begin{equation}\label{eq:ha}
  \begin{split}
    \h{A} & = H_\mathrm{loc} + H_\mathrm{bath} \\
    H_\mathrm{loc} & = \sum_{\{\tau\}} \epsilon_{\tau_1\tau_2} a^\dag_{\tau_1} a_{\tau_2} + \sum_{\{\tau\}} U_{\tau_1\tau_2\tau_3\tau_4} a^\dag_{\tau_1} a^\dag_{\tau_2} a_{\tau_4} a_{\tau_3}\\
    H_\mathrm{bath} & = \sum_{\kappa} \epsilon_{\kappa} a^\dag_\kappa a_\kappa + \sum_{\tau,\kappa} V_{\tau\kappa} a^\dag_\tau a_\kappa + \mathrm{H.c.}
  \end{split}
\end{equation}
where a locally interacting impurity site ($H_\mathrm{loc}$) is coupled to a non-interacting bath ($H_\mathrm{bath}$). The fermionic operators $a^{(\dag)}_{\tau/\kappa}$ annihilate (create) an electron labeled by a set of quantum numbers $\tau$ or $\kappa$ on the impurity site $i$ or bath sites $l$. In addition to the Coulomb interaction terms, the local Hamiltonian $H_\mathrm{loc}$ typically includes single-particle operators such as crystal-field (dominant in 3$d$ electron systems) and spin-orbit coupling (relevant in $4/5d$ and $4f$ systems). $H_\mathrm{bath}$ includes the bath dispersion and its coupling to the impurity. The Hamiltonian~\eqref{eq:ha} is depicted in Fig.~\ref{fig:geo}(a). In this section, we present a natural-orbital representation of the impurity model, which can be combined with a projection scheme to efficiently obtain the ground state and one-body Green's function on the real-frequency axis.

\subsection{Natural-Orbital Representation of an Impurity Model}

The \emph{natural orbitals} are defined as a single-particle basis set on which the ground-state single-particle density matrix of a quantum system is diagonal. They are widely used in quantum chemistry~\cite{Wilson2014book} as they have several advantageous features for molecular systems such as optimal convergence properties for the wave functions and energies. In the context of quantum impurity problems, they have been discussed in conjunction with configuration-interaction expansion approximation~\cite{Zgid2012,Lin2013}. These methods have proven to be capable of solving impurity problems exceeding the size of those dealt by conventional ED~\cite{Caffarel1994}. The caveat of employing natural orbitals for impurity models is that a naive implementation that diagonalizes the density matrix of the \emph{whole} system inevitably mixes the impurity states with the non-interacting bath states. This transforms the original local interactions in $H_\mathrm{loc}$ into long range ones in the resultant Hamiltonian, which may bring a severe penalty that overcomes the advantage of the natural orbitals, especially for large systems that contains $\mathcal{O}(10^2)$ bath sites.

In Ref.~\cite{Lu2014}, some of us have introduced a natural-orbital representation of the impurity model by restricting the optimization of the basis set only for the bath degrees of freedom. It was shown that an ED solver employing such a natural-orbital basis set substantially outperforms conventional ones and is capable of solving impurity models with the number of bath sites comparable to that achieved by NRG or DMRG solvers~\cite{Lu2014}. The resulting geometry of the impurity Hamiltonian is graphically represented in Fig.~\ref{fig:geo}(b). The procedure for obtaining such a representation is detailed in Ref.~\cite{Lu2014}. We briefly recapitulate the steps here:
\begin{enumerate}[label=(\roman*)]
  \item Solve Hamiltonian~\eqref{eq:ha} (as depicted in Fig.~\ref{fig:geo}(a)) within mean-field methods (e.g. Hartree-Fock) and obtain the ground-state single-particle density matrix $\hat \rho^\mathrm{MF}=\bigl( \begin{smallmatrix} \hat \rho_i & \hat \rho_{il}\\ \hat \rho_{li} & \hat \rho_{l} \end{smallmatrix}\bigr)^\mathrm{MF}$, where we distinguish the impurity ($i$) and bath ($l$) parts explicitly.
  \item Diagonalize the bath density matrix $\hat \rho_{l}^\mathrm{MF}$, which leads to a new set of bath orbitals with occupation of either 0 or 1, with the exception of $m$ (the number of impurity spin-orbitals) orbitals that have fractional occupation. We assign these orbitals to site $b$ as shown in Fig.~\ref{fig:geo}(b). Its density matrix $\hat \rho_b^\mathrm{MF}$ satisfies the relation $\Tr \hat \rho_b^\mathrm{MF} = m - \Tr \hat \rho_i^\mathrm{MF}$.
  \item Linearly combine the impurity site $i$ and bath site $b$ into ``bonding'' and ``anti-bonding'' sites with occupation $m$ and $0$. The former (latter) only couples to the completely filled (empty) bath sites obtained from last step, respectively. The mean-field Hamiltonian has now been separated into two decoupled terms, each describes the filled or empty spin-orbitals of the complete single-particle Hilbert space.
  \item Perform unitary transformation (Lanczos tridiagonalization) on the two parts of the Hamiltonian and obtain two separate empty and filled ``chains'' starting with the bonding and anti-bonding sites, respectively.
  \item Finally, reverse the unitary transformation in step (iii) and recover the $i$ and $b$ sites, which now couple to both the empty and filled chains. Following the convention in Ref.~\cite{Lu2014}, we dub the two chains ``conduction'' and ``valence'' baths, respectively.
\end{enumerate}

In the limit of $U \rightarrow 0$, these mean-field natural orbitals are exact, and the many-body ground state of the exact impurity solution can be written out using only $2m$ Slater determinants~\cite{Lu2014}. At finite $U$ values, the exact occupation of the conduction or valence bath sites will deviate from 0 or 1, necessitating the inclusion of more states with excited electrons or holes in the conduction or valence chains. Nonetheless, the ``leakage'' of electrons (holes) onto a conduction (valence) site is expected to rapidly decay as a function of its distance to the impurity site, as states with electrons (holes) deep in the conduction (valence) chain are energetically unfavorable. This allows for an efficient description of the ground state and the low-energy excitations by only including states with electron (hole) excitations in the conduction (valence) bath that are localized around the impurity site.

\begin{figure}[tb]
  \includegraphics{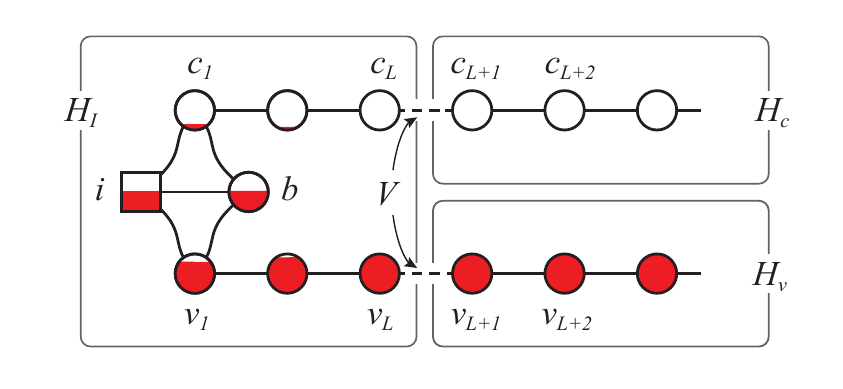}
  \caption{\label{fig:proj} Separating the full Anderson impurity Hamiltonian $\h{A}$ defined on the natural orbitals into four parts $\h{I}$, $\h{c}$, $\h{v}$, and $V$. Each Hamiltonian acts on the sites enclosed by its corresponding box. The hybridization operator $V$ (dashed bonds) connects $\h{I}$ to $\h{c}$ and $\h{v}$.}
\end{figure}

\subsection{Ground-State Projection}\label{sec:proj:gs}

So far we have rewritten the impurity Hamiltonian~\eqref{eq:ha} on the natural-orbital basis, which was shown to be a highly efficient representation of the impurity model. Such a representation has an optimal scaling behavior with respect to the number of bath sites, as adding empty (filled) bath sites at the end of the conduction (valence) chains incurs little to none cost for describing the ground state. However, the computation complexity is still expected to scale exponentially with the number of impurity spin-orbitals, which, depending on the occupation and the exact form of the Hamiltonian, may become intractable for full $d$/$f$-orbital impurities that are each coupled to a few hundred bath sites.

To further reduce the computation cost and alleviate the scaling problem, for the ground state, we follow ideas from a restrictive active space calculation, similar to the optimizations made by Gunnarsson and Sch\"onhammer~\cite{Gunnarsson:1983uf} for the calculations of an $f$-level Anderson impurity model for Ce compounds. These methods are currently often used for ligand-field theory calculations for core-level spectroscopy~\cite{Haverkort2012}. We propose to project the full Hilbert space onto a subspace that only contains states with completely empty conduction (filled valence) sites with indices $l > L$ (Fig.~\ref{fig:proj}), with $L$ as a tunable parameter controlling the trade-off between projection accuracy and computation cost. Note that a single $L$ is used here for simplicity. For a general multi-orbital impurity model, $L$ does not need to be the same for the conduction and valence bath or for different spin-orbitals. The projection essentially separates the full Hamiltonian $\h{A}$ into three parts: an impurity Hamiltonian $\h{I}$ of a much smaller system, as well as $\h{c}$ and $\h{v}$ describing two truncated bath chains that are coupled to $\h{I}$ via hybridization $V$. The Anderson impurity Hamiltonian is then
\begin{equation}
  \h{A} = \h{0} + V,
\end{equation}
where
\begin{equation}
  \h{0} = \h{I} + \h{v} + \h{c}.
 \end{equation}
The projected ground state wave function is given as
\begin{equation}
  \ket{\Phi_0} = \ket{\phi_\mathrm{I}}\otimes\ket{\mathbbm{0}_c}\otimes\ket{\mathbbm{1}_v}.
\end{equation}
Here, $\ket{\phi_\mathrm{I}}$ is the exact ground state of $\h{I}$ that can be efficiently computed by ED or DMRG methods for moderately large $L$, and $\ket{\mathbbm{0}_v}$ ($\ket{\mathbbm{1}_c}$) denotes the product states of completely empty conduction (filled valence) sites with indices $l > L$. $\ket{\Phi_0}$ is the exact solution of $\h{0}$. The projected ground-state energy is
\begin{equation}\label{eq:gse}
    E_0(L) \equiv \ev{\h{A}}{\Phi_0} = \ev{\h{I}}{\phi_\mathrm{I}} + \sum_{l > L, m} \epsilon^v_{lm},
\end{equation}
where the second term is simply the sum of on-site energies of all spin-orbitals with indices $m$ at each site $l$ in the truncated valence chain. The accuracy of the projected ground-state wave function can be assessed by calculating the deviation of Eq.~\eqref{eq:gse} from the exact ground-state energy when the latter is attainable, or by calculating the energy variance of the projected ground-state using the full Hamiltonian $\h{A}$ as
\begin{equation}\label{eq:variance}
  \begin{split}
    \delta E_0(L)^2 & \equiv \ev{\h{A}^2}{\Phi_0}-E_0(L)^2 \\
    & = \delta E_\mathrm{I}(L)^2 + \delta V(L)^2,
  \end{split}
\end{equation}
with
 \begin{equation*}
  \begin{split}
    \delta E_\mathrm{I}(L)^2 & = [\ev{\h{I}^2}{\phi_\mathrm{I}}-\ev{\h{I}}{\phi_\mathrm{I}}^2 ] \\
    \delta V(L)^2 & = \ev{V^2}{\Phi_0}.
  \end{split}
\end{equation*}
The first term $\delta E_\mathrm{I}(L)^2$ is intrinsic to the numerical method of choice that solves $\h{I}$. The second term $\delta V(L)^2$ originates from the imposed projection and therefore scales exponentially to zero with increasing $L$.

\subsection{Excited-State Projection and Green's Functions}\label{sec:proj:gf}

The central object of interest for an impurity problem is the impurity Green's function $G_\mathrm{imp}(\omega)$. On the \emph{real-frequency} axis, it is defined as
\begin{equation}
  G_\mathrm{imp}(\omega) = G^+(\omega) - G^-(-\omega)^\ast,
\end{equation}
where $G^{\pm}(\omega)$ are the retarded Green's functions for electron addition ($+$) and removal ($-$) at the impurity site $i$:
\begin{equation}
  \begin{split}
    G^+(\omega) & = \lim_{\eta\rightarrow 0^+} \ev{ a_i \frac{1}{\omega-\h{A}+ \ii \eta} a^\dag_i }{\Psi_0} \\
    G^-(\omega) & = \lim_{\eta\rightarrow 0^+} \ev{ a^\dag_i \frac{1}{\omega-\h{A}+ \ii \eta} a_i }{\Psi_0},
  \end{split}
\end{equation}
with $\ket{\Psi_0}$ the impurity ground state. The Green's functions can be directly calculated in the frequency domain using Lanczos method, which is an approach generally adopted in ED-based solvers~\cite{Caffarel1994,Sangiovanni2006,Capone2007,Koch2008,Zgid2012,Lin2013,Lu2014}. For DMRG solvers, $G_\mathrm{imp}(\omega)$ is also commonly obtained via Fourier transform from the real-time Green's functions~\cite{White2004,Ganahl2015}. While the proposed projection scheme is applicable for both methods, in this paper, we will focus on the direct calculation in the frequency domain.

The idea of our projection method is to obtain the impurity Green's function of the full system $G_\mathrm{imp}(\omega)$ from that of the projected system $G_0(\omega)$ given by $\h{0}$ and successive non-perturbative expansion in the hybridization $V$. Such an expansion can in principle be done using diagrammatic methods and the Dyson equations. This requires knowledge not only on the impurity Green's function of $\h{0}$, but also on electron (hole) propagators starting at site $c_L$ ($v_L$). Here, however, we employ a method based on Hilbert space reductions, which has the advantage that we can use standard Lanczos routines for solving the Green's functions of impurity models.

The method is based on the notion that we can connect to each operator $H$ with a fixed number of electrons a Hilbert space $\mathcal{H}$. We start with the projected subspace $\mathcal{H}_0 = \mathcal{H}_\mathrm{I} \otimes \ket{\mathbbm{1}_v}\otimes \ket{\mathbbm{0}_c}$ defined for the ground-state calculation, where $\mathcal{H}_\mathrm{I}$ is the Hilbert space of the subsystem $\h{I}$. To obtain $G^\pm_0(\omega)$, we use the Lanczos method and construct a series of $M$ Krylov vectors $\ket{\tilde \nu_j}=\h{I}^j a_i^{(\dag)} \ket{\phi_\mathrm{I}} \in \mathcal{H}^\prime_\mathrm{I}$, where the prime denotes the Fock subspaces of electron removal (addition) with respect to $\mathcal{H}_\mathrm{I}$. After orthogonalizing each $\ket{\tilde \nu_j}$ to the previous states and proper normalization, the resultant set of vectors $\{\ket{\nu_j}\}$ become the basis set of a subspace (Krylov space) $\mathcal{K}^M$ of $\mathcal{H}^\prime_\mathrm{I}$ with dimension $M$. The Hamiltonian $\h{I}$ is represented as a tridiagonal matrix $\overbar H_\mathrm{I}$ on $\mathcal{K}^M$, and $G^\pm_0(\omega)$ can be straightforwardly calculated as the leading element of the resolvents $G^{\pm}_0(\omega) = (\omega + \ii \eta - \overbar H_\mathrm{I})^{-1}_{00}$, which is conveniently expressed as a continued fraction~\cite{Lu2014}. The corresponding impurity Green's functions $G^\pm_0(\omega)$ are identical to those of the subsystem $\h{I}$.

The $G_0^\pm(\omega)$ obtained above are in general quite different from the Green's functions $G^\pm(\omega)$ of the full system, especially for small $L$ values, due to the limited degrees of freedom. To obtain a more accurate description, we need to relax the projection condition to include more excited states. This can be done by allowing electron (hole) excitations into the completely empty conduction (filled valence) chains. As states with higher-order excitations are energetically more costly and therefore contribute less to the Green's functions, the number of excited particles $p$ serves as a control parameter for the projection. Conceptually this is similar to the restricted active space method used in quantum chemistry.

The proposed projection scheme can be implemented in ED and DMRG solvers by targeting a specific $U(1)$ symmetry sector for the bath chains in each step of the Lanczos or time-evolution process when computing the Green's function. Here, we combine it with further simplification by manually identifying the relevant states for $p$-particle excitations. While it might seem cumbersome at first, the advantage of such a procedure is that it allows for the calculation of the \emph{full} Green's function by evaluating Hamiltonian matrix elements on the basis of $\mathcal{K}^M$ and their derived states with singly ($p=1$) and doubly ($p=2$) excited particles in the bath chains. This essentially reduces the solution of a many-body problem $\h{A}$ with a few hundred spin-orbitals to that of the much smaller subsystem $\h{I}$.

\subsubsection{$p=1$ projection}
In the following, we derive the expression of the Hamiltonian and the Green's functions on the expanded subspace that includes single-electron (hole) excitations into the conduction (valence) chain. For simplicity, we assume a single-orbital model, as the generalization to multi-orbital case is straightforward. We further omit spin indices as the expressions are spin independent.

The expanded states with single-particle excitations in the bath chains can be obtained by acting $\h{A}$ on the initial subspace $\mathcal{H}^\prime_0=\mathcal{H}^\prime_\mathrm{I} \otimes \ket{\mathbbm{1}_v}\otimes \ket{\mathbbm{0}_c}$. As $\mathcal{H}^\prime_0$ is closed under $H_0$, the singly excited states are then generated by $V\mathcal{H}^\prime_0$. Note that $\mathcal{H}^\prime_I$ is still exponentially large for a sufficiently large $L$, in practice we approximate it by $\mathcal{K}^M$, which is known to provide an accurate representation for $\h{I}$. $\mathcal{H}^\prime_0$ is then replaced by $\overbar {\mathcal{H}}^\prime_0 = \mathrm{span}(\ket{\psi_j} = \ket{\nu_j}\otimes\ket{\mathbbm{1}_v}\otimes\ket{\mathbbm{0}_c}| j=0,\dots,K)$. The $p=1$ expanded vector space $\mathcal{H}^\prime_1$ is therefore approximately given as
\begin{equation*}
  \mathcal{H}^\prime_1 \approx \overbar {\mathcal{H}}^\prime_1 = \mathrm{span}(\{\ket{\psi^e_{jk}}\}) + \text{span}(\{\ket{\psi^h_{jk}}\}),
\end{equation*}
where
\begin{equation*}
  \begin{split}
    \ket{\psi^e_{jk}} & = t_c c_{L}\ket{\nu_j}\otimes\ket{e_k} = \ket{\eta_j}\otimes\ket{e_k},\quad \text{and} \\
  \ket{\psi^h_{jk}} & = t_v v^\dag_{L}\ket{\nu_j}\otimes\ket{h_k} = \ket{\zeta_j}\otimes\ket{h_k},
  \end{split}
\end{equation*}
where $\ket{e_k}=\ket{\mathbbm{1}_v}\otimes c^\dag_{L+k}\ket{\mathbbm{0}_c}$ and $\ket{h_k}=v_{L+k}\ket{\mathbbm{1}_v}\otimes\ket{\mathbbm{0}_c}$ ($k\geq 1$) are the single electron and hole states of the truncated bath chains. We have relabeled the fermionic operators on the conduction and valence sites by $c^{(\dag)}$ and $v^{(\dag)}$, respectively. $V$ is now explicitly given as $V=t_c c^\dag_{L} c_{L+1} + t_v v^\dag_{L} v_{L+1} +\mathrm{H.c.}$, where $t_{c(v)}$ is the hopping between conduction (valence) bath sites $L$ and $L+1$ (see Fig.~\ref{fig:proj}). It is easily seen that $\overbar {\mathcal{H}}^\prime_1$ is orthogonal to $\overbar {\mathcal{H}}^\prime_0$. We can evaluate the matrix elements of $\h{A} = \h{0} + V$ on the $p\leq 1$ subspace as
\begin{equation}\label{eq:V1}
  \begin{split}
    \bra{\psi_j}H_0\!\ket{\psi_k} &= \mel{\nu_j}{H_I}{\nu_k} = H^I_{jk} \\
    \bra*{\psi^{e}_{jk}}H_0\!\ket*{\psi^{e}_{lm}} & = \mel*{\eta_j}{H_I}{\eta_l} \updelta_{km} + \mel{e_k}{H_c}{e_m} \updelta_{jl} \\
    & = H^{I\eta}_{jl} \updelta_{km} + H^c_{k,m} \updelta_{jl} \\
    \bra*{\psi^{h}_{jk}}H_0\!\ket*{\psi^{h}_{lm}} & = \mel*{\zeta_j}{H_I}{\zeta_l} \updelta_{km} + \mel{h_k}{H_v}{h_m} \updelta_{jl} \\
    & = H^{I\eta}_{jl} \updelta_{km} + H^v_{km} \updelta_{jl}. \\
\end{split}
\end{equation}
and
\begin{equation}\label{eq:V1}
  \begin{split}
    \bra{\psi_j}V\!\ket{\psi^e_{kl}} & = \braket{\eta_j}{\eta_k} \updelta_{0l} = V^{\eta}_{jk}\updelta_{0l}\\
    \bra{\psi_j}V\!\ket{\psi^h_{kl}} &= \braket{\zeta_j}{\zeta_k} \updelta_{0l} = V^{\zeta}_{jk}\updelta_{0l}.
\end{split}
\end{equation}
Note that we have defined the ground-state energy to be zero. The elements of the matrices $H^I(\equiv \overbar{H}_I)$, $H^{c}$, and $H^{v}$ are already known. One only needs to evaluate the ($M$-dimensional) matrices $H^{I\gamma}$ and $V^{\gamma}$ ($\gamma = \eta,\,\zeta$), with the latter identified with the overlap matrix of $\{\ket{\gamma}\}$. Note that the states $\{\ket*{\psi^{e(h)}_{jk}}\}$ are not orthonormal. To bring them into an orthonormal form, one can solve the generalized eigenvalue problem $H^{I\gamma}$ with respect to $V^{\gamma}$ and obtain the eigenvector matrix $T^{\gamma}$. The above matrices are then expressed on the orthonormal basis set as $\tilde H^{I\gamma} = {T^{\gamma}}^\dag H^{I\gamma} T^{\gamma}$, which is the diagonal eigenvalue matrix, and $\tilde V^{\gamma} = V^{\gamma} T^{\gamma}$. The Green's function can then be calculated by inverting the full $\h{A}$ defined on $\overbar{\mathcal{H}}^1$.

\subsubsection{$p=2$ projection}
We further relax the projection condition to allow double excitations. The full $p\leq 2$ subspace is given by $\h{A}^2 \mathcal{H}^\prime_0 = \mathcal{H}^\prime_0 + \mathcal{H}^\prime_1 + V^2\mathcal{H}^\prime_0$. The $p=2$ subspace $\mathcal{H}^\prime_2$ is then spanned by the subset of doubly excited states in $V^2\mathcal{H}^\prime_0 = V\mathcal{H}^\prime_1$. Similar to the $p=1$ case, we approximate $\mathcal{H}^\prime_2\approx\overbar{\mathcal{H}}^\prime_2=V\overbar{\mathcal{H}}^\prime_1$~\footnote{We note that this effectively remove states such as $VH^i\mathcal{H}^\prime_1$ ($i \in \mathbb{N}$) from the complete $p=2$ subspace, which are found to be heavily overlapping with those already contained in $V\overbar{\mathcal{H}}^\prime_1$. They will thus be purged in the subsequent orthogonalization process and do not expand the Hilbert space}. Under such approximation, the states in $\overbar{\mathcal{H}}^\prime_2$ are given as
\begin{equation*}
  \begin{split}
  \psi^{e\uparrow e\downarrow}_{jkl} & = t_c^2 c_{L,\uparrow} c_{L,\downarrow} \ket{\nu_j} \!\otimes\! \ket{e_{k\uparrow}e_{l\downarrow}} = \ket{\lambda_j} \!\otimes\! \ket{e_{k\uparrow}e_{l\downarrow}} \\
  \psi^{h\uparrow h\downarrow}_{jkl} & = t_v^2 v^\dag_{L,\uparrow} v^\dag_{L,\downarrow} \ket{\nu_j} \!\otimes\! \ket{h_{k\uparrow}h_{l\downarrow}} = \ket{\mu_j} \!\otimes\! \ket{h_{k\uparrow}h_{l\downarrow}} \\
  \psi^{e\sigma h\sigma'}_{jkl}\!\!\! & = t_c t_v c_{L,\sigma} v^\dag_{L,\sigma'}\ket{\nu_j} \!\otimes\! \ket{e_{k\sigma}h_{l\sigma'}} = \ket{\theta_j} \!\otimes\! \ket{e_{k\sigma}h_{l\sigma'}}
  \end{split}
\end{equation*}
which describe two-electron, two-hole, and electron-hole excitations into the bath chains. The spin indices are recovered here considering the Pauli principle. The Hamiltonian matrix elements for $H_0$ read
\begin{equation}
  \begin{split}
    \bra*{\psi^{e\uparrow e\downarrow}_{jkl}\!}&H_0\!\ket*{\psi^{e\uparrow e\downarrow}_{mno}\!} \!=\! H^{I\lambda}_{jm} \updelta_{kn} \updelta_{lo} \!+\! H^{c}_{kn} \updelta_{jm} \updelta_{lo} \!+\! H^{c}_{lo} \updelta_{jm} \updelta_{kn} \\
    \bra*{\psi^{e\sigma h\sigma'}_{jkl}\!}&H_0\!\ket*{\psi^{e\sigma h\sigma'}_{mno}\!} \!=\! H^{I\theta}_{jm} \updelta_{kn} \updelta_{lo} \!+\! H^{c}_{kn} \updelta_{jm} \updelta_{lo} \!+\! H^{v}_{lo} \updelta_{jm} \updelta_{kn},
  \end{split}
\end{equation}
with $H^{I\lambda}$ and $H^{I\theta}$ the Hamiltonian matrices evaluated on the basis set $\{\ket{\lambda}\}$ and $\{\ket{\theta}\}$. The matrix elements for $V$ are given as
\begin{equation}
  \begin{split}
    \bra*{\psi^{e\uparrow}_{jk}}&V\!\ket*{\psi^{e\uparrow e\downarrow}_{lmn}} = \braket{\theta_j}{\theta_l} \updelta_{k,m} \updelta_{0,n} = V^{\theta}_{jl} \updelta_{k,m} \updelta_{0,n} \\
    \bra*{\psi^{e\downarrow}_{jk}}&V\!\ket*{\psi^{e\uparrow e\downarrow}_{lmn}} = \braket{\theta_j}{\theta_l} \updelta_{k,m} \updelta_{0,n} = V^{\theta}_{jl} \updelta_{0,m} \updelta_{k,n}.
  \end{split}
\end{equation}
Similar expressions can also be derived for the excited-hole states. Same as for the $p=1$ case, the expanded states $\{\ket{\lambda}\}$, $\{\ket{\mu}\}$, $\{\ket{\theta}\}$ need to be orthonormalized.

Finally, we emphasize that as we approximate the electron addition/removal Hilbert space $\mathcal{H}^\prime_0$ of the subsystem $\h{I}$ by $\mathcal{K}^M$, the completeness of the $p=1$ and $p=2$ states depends on $M$, and the results should be tested for convergence in $M$.

\section{DMFT}

We now demonstrate an application of the natural-orbital solver presented above in the context of DMFT~\cite{Metzner1989,Georges1996}. Within DMFT, a Hubbard model is mapped onto a single-impurity Anderson model supplemented by a self-consistency condition that identifies the impurity Green's function with the local lattice one. The central ingredient of DMFT is thus the (iterative) calculation of the impurity Green's function. The steps for constructing the DMFT self-consistency loop entirely on the real-frequency for a general Hamiltonian can be found in e.g. Ref.~\cite{Lu2014, Bauernfeind2017}. It should be noted that the prerequisite of such constructions is to include a sufficiently large number ($\mathcal{O}(10^2)$) of bath sites in the Hamiltonian~\eqref{eq:ha}, which is necessary for an accurate real-frequency representation of the bath Green's function. This also guarantees that the computed self energy in a general DMFT loop is always causal~\cite{Lu2014}, which could otherwise be an issue for conventional real-frequency (configuration-interaction) implementations that include only a limited number of bath sites.

In the following sections, we focus our discussion on the calculation of the one-band Hubbard model on the Bethe lattice with infinite coordination number, for which the DMFT mapping is exact. The corresponding impurity Hamiltonian is given as
\begin{equation}\label{eq:ha1}
  \begin{split}
    H_A = & \sum_\sigma \epsilon_i n_{i\sigma} + U n_{i\uparrow} n_{i\downarrow} + \\
  & \sum_{l\sigma} \epsilon_l n_{l\sigma} + \sum_{l\sigma} (V_l a^\dag_{i\sigma} a_{l\sigma} + \mathrm{h.c.}),
  \end{split}
\end{equation}
where $i$ and $l$ denote the impurity and bath sites, respectively. In addition, for benchmark purposes we assume spin-symmetric couplings and particle-hole symmetry, as there is abundant literature containing high quality results obtained from different numerical methods. In this case, the the DMFT loop can be greatly simplified, as the imaginary part of the bath hybridization function $\tilde \Delta(\omega)\equiv-\frac{1}{\pi}\Im \Delta(\omega) = \sum_l \abs{V_l}^2 \updelta (\omega-\epsilon_l)$ is related to the impurity spectral function $A_\mathrm{imp}(\omega)\equiv-\frac{1}{\pi}\Im G_\mathrm{imp}(\omega)$ as $\tilde \Delta(\omega) = \frac{D^2}{4} A_\mathrm{imp} (\omega)$, where $D$ is the half-bandwidth of the semi-elliptic noninteracting density of states. The spin indices for the observables are omitted hereafter for the ease of notation.

Within each DMFT loop, the bath parameters are obtained by a discrete representation of the  hybridization function over $N_l\sim\mathcal{O}(10^2)$ poles. We employ a scheme similar to Ref.~\cite{Bulla2005} by discretizing the frequency-axis into $N_l$ intervals \{$I_l$\}, and obtain $V_l$ and $\epsilon_l$ as
\begin{equation}\label{eq:dis}
  \begin{split}
    V_l^2 & = \int_{I_l} \dd{\omega} \tilde \Delta(\omega), \\
    \epsilon_l & = \frac{1}{V_l^2} \int_{I_l} \dd{\omega} \omega \tilde \Delta(\omega).
  \end{split}
\end{equation}
Here we chose the intervals such that the weight $V_l^2$ is equal for each bath site. We note that the details of the discretization scheme has little effect on the results when the number of bath sites is large enough. The impurity spectral function $A_\mathrm{imp}(\omega)$ is then obtained by solving the resulting impurity model with our solver described in Sec.~\ref{sec:solver}, which leads to an update of the hybridization function
\begin{equation}
  \tilde \Delta(\omega) = \frac{D^2}{4} \left[\alpha A^\prime_\mathrm{imp}(\omega) + (1-\alpha)A_\mathrm{imp}(\omega)\right],
\end{equation}
with $\alpha \in [0,1)$ a mixing factor that allows for under-relaxation by mixing in the spectral function $A^\prime_\mathrm{imp}(\omega)$ from the previous loop. The convergence is reached once $A_\mathrm{imp}(\omega)=A^\prime_\mathrm{imp}(\omega)$.

\section{Results}

We note that while the natural-orbital representation and projection scheme in Sec.~\ref{sec:solver} can be readily implemented in existing ED solvers~\cite{Lu2014}, we adopt the MPS-based DMRG method~\cite{Schollwock2011} here for computing the impurity ground state and Green's functions, which is expected to be more efficient for large $L$ values considering the quasi one-dimensional geometry in Fig.~\ref{fig:geo}(b). We use the zip-up method when multiplying a Hamiltonian (as a matrix-product operator) to MPS~\cite{Stoudenmire2010} for generating the Krylov states. Note that due to the relatively small size of the subsystem $\h{I}$, the total truncated weight of the MPS in each Lanczos step can be kept well below $10^{-16}$.

In the following, we present DMFT results obtained for the one-band Hubbard model on the Bethe lattice using the proposed projection method. The total number of bath sites is set to $N_l=301$, with each bath chain of full length 150. With such a setting the coexisting region of the metallic and insulating solution is found between $U/D=$ 2.40 and 3.10, in close agreement with previous results obtained using NRG~\cite{Bulla2011}. The presented calculation is performed for interaction values $U/D$ ranging from $1/16$ to $16$, including both the itinerant and atomic limits. Especially, we focus our discussion on three representative values $U/D=1.0$, $2.0$, and $4.0$, which correspond to weakly-correlated metal, strongly-correlated metal, and Mott insulator ground states in DMFT, respectively~\cite{Georges1996}.

\subsection{Ground State Convergence}

\begin{figure}[tb]
  \includegraphics{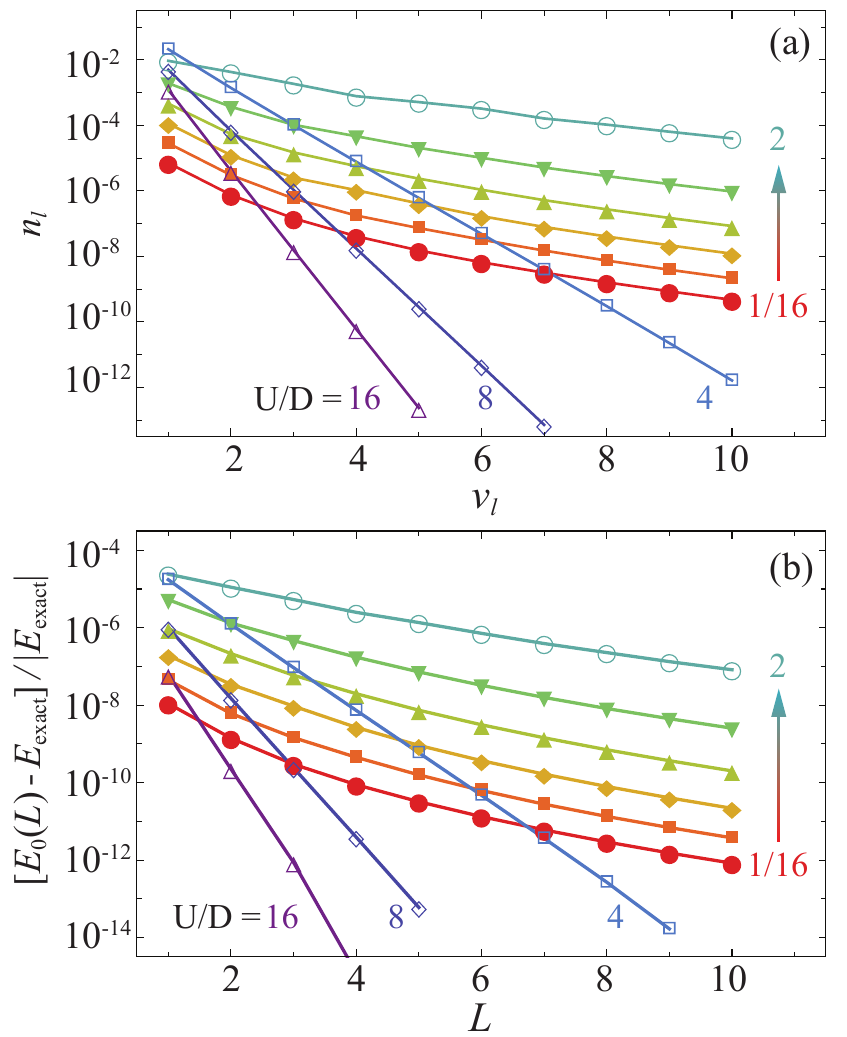}
  \caption{\label{fig:dE} (a) Number of electrons per spin-orbital on the first 10 conduction bath sites in the converged DMFT ground state for $U/D$ values ranging from $1/16$ to $16$ as a geometric sequence with common ratio 2.  The values are noted next to each curve. (b) Normalized ground-state energy deviation $[E_0(L)-E_\mathrm{exact}]/\abs{E_\mathrm{exact}}$ as a function of $L$.}
\end{figure}

\begin{figure*}[tb]
  \includegraphics{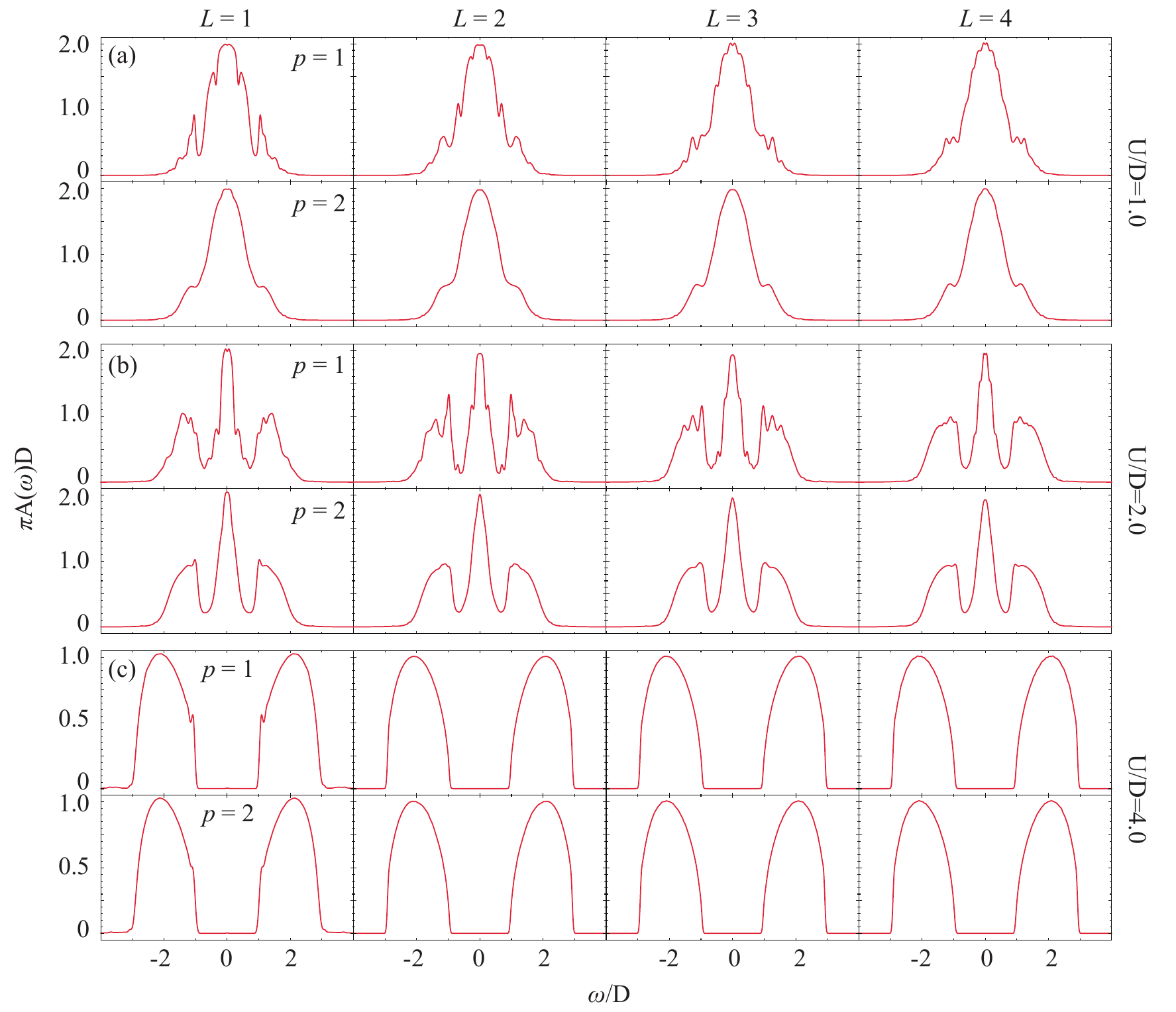}
  \caption{\label{fig:sf} DMFT spectral functions for (a) $U/D=1.0$, (b) $U/D=2.0$, and (c) $U/D=4.0$ calculated with different projection parameters $(L, p)$.}
\end{figure*}

We start by discussing the ground-state results for the different $U$ values. Fig.~\ref{fig:dE}(a) shows the number of electrons per spin-orbital on the first 10 conduction bath sites in the converged DMFT ground state. Note that this is identical to the hole occupation in the valence chain due to the particle-hole symmetry. In the metallic regime with $U/D$ from $1/16$ to $2$, $n_l$ on each site converges towards 0 with decreasing $U$ values. This is expected as the natural orbitals are exact in the $U \rightarrow 0$ limit. For a given $U$, we observe near-exponential decay of $n_l$ with increasing site index $l$. Exact exponential decay of $n_l$ is observed for the insulating cases with $U/D \geq 4$, as any particle-hole excitations into the bath chains is suppressed by the Mott gap of approximately $U-2D$. The slowest convergence is observed for the correlated metals with $U/D \sim 2$, yet the electron density reaches below $10^{-3}$ within the first two to four bath sites for all the cases considered here. Closer inspection of the ground-state wave function reveals that even for the worst cases, states with completely empty conduction (filled valence) bath sites for $l \geq 2$ comprise more than 99\% of the total weight, which justifies the proposed $p=0$ projected wave function in Sec.~\ref{sec:proj:gs} as a valid approximation for the exact ground state.

Fig.~\ref{fig:dE}(b) shows the relative error of the projected ground-state energy $E_0(L)$ (Eq.~\eqref{eq:gse}) when applying projection at bond $L$ between bath sites $L$ and $L+1$ (see Fig.~\ref{fig:proj}). As the energy deviation is directly correlated with the ground-state electron (hole) density in the conduction (valence) chains, one observes that $E_0(L)$ converges exponentially to the exact DMRG ground-state energy for the full system $E_\mathrm{exact}$.

\subsection{Green's Functions}

We proceed to calculate the DMFT Green's functions with a few different sets of control parameters $(L, p)$. The calculated spectral functions are presented in Fig.~\ref{fig:sf}(a)--(c) for $U/D=$ 1.0, 2.0, and 4.0, respectively. The spectra are convoluted with a Gaussian kernel with full width at half maximum of $0.04D$.

The first row of each panel shows the spectral functions calculated with $p=1$. Within each row, the results are presented for $L=1$ on the left up to $L=4$ on the right. For all $U$ values, the spectra retain the general line shape of previous results~\cite{Lu2014,Ganahl2014,Ganahl2015}. This is best seen for the $U/D=2.0$ case in Fig.~\ref{fig:sf}(b), where the spectra show a sharp resonance at $\omega=0$ and two broad Hubbard bands at approximately $\omega=\pm U/2$. Especially, the Luttinger pinning~\cite{MH1989} at $\omega=0$ with the condition $\pi D A(\omega=0)=2.0$ is fulfilled to a high accuracy for the metallic cases in Fig.~\ref{fig:sf}(a) and (b). This suggests that the $p=1$ projected states, i.e. those with only single-particle excitations in the bath sites, indeed capture the low-energy physics of the impurity model. On the other hand, we notice spurious oscillatory features/small peaks on the side of the quasiparticle peak or on the Hubbard bands, most noticeably for the metallic cases. As the amplitude of these features decreases with increasing $L$, they can be attributed partially to the missing of states with multi-particle excitations in bath sites close to the impurity site in the $p=1$ projected subspace. We also note that for the insulating case in Fig.~\ref{fig:sf}(c), there is some small residual weight (smaller than $10^{-4}$) close to $\omega=0$ for $L=1$, which vanishes for $L\geq2$.

The second row of each panel shows the spectral functions calculated with $p=2$. Compared to the $p=1$ results, the oscillatory features are greatly suppressed and smooth spectra are recovered for all $U$ and $(L,p)$ values. The results for $L=3$ and 4 are in excellent agreement with previous results obtained using time evolving block decimation (TEBD)~\cite{Ganahl2015} (see Appendix). For the case of $U/D=2.0$, two sharp side peaks can be observed at the inner edges of the Hubbard bands, in line with previous ED or DMRG results~\cite{Lu2014,Ganahl2014,Ganahl2015,Wolf2014}. We do note that the exact size of the side peaks is $L$ dependent, and shows a converging behavior with increasing $L$ similar to that of a Fourier spectral decomposition with increasing frequency cutoff. It is also closely related to the observation in Ref.~\cite{Ganahl2015}, where the peak position and size are dependent on the system size (number of bath sites), and are likely related to the time-dependent probability of the impurity being doubly occupied. For the insulating case, we note that the change of $A(\omega)$ between the $p=1$ and $p=2$ results is less than $10^{-3}$ at all frequencies for $L\geq2$.

\begin{figure}[tb]
  \includegraphics{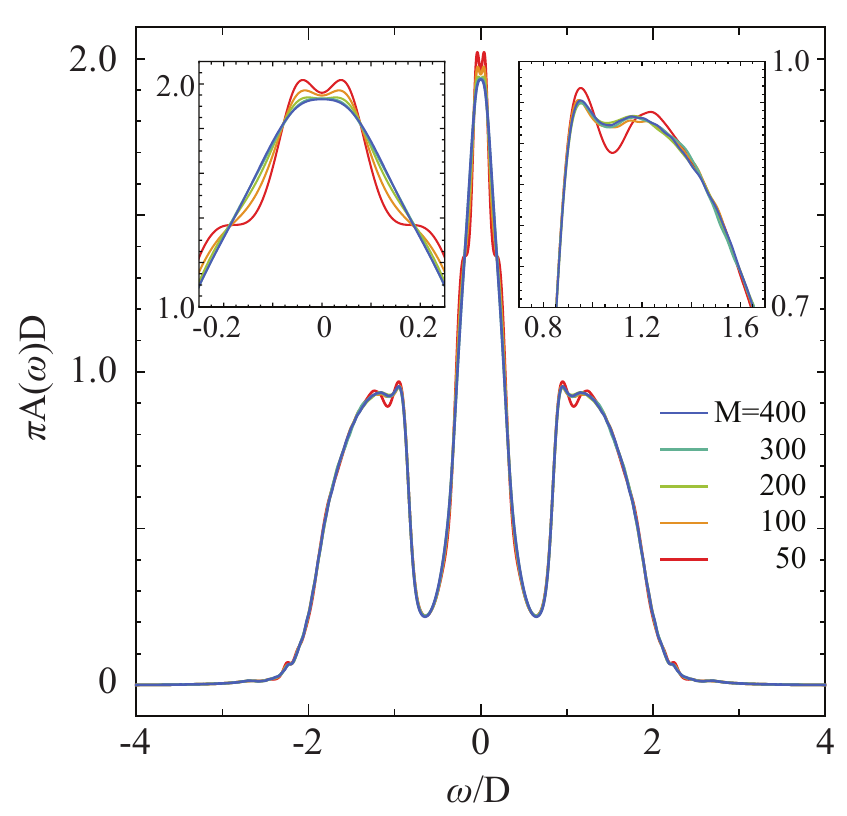}
  \caption{\label{fig:convergeK} DMFT spectral function for $U/D=2.0$ with different sizes of the initial Krylov space $M$. The insets show the detail of the quasiparticle peak and the Hubbard band.}
\end{figure}

As mentioned before, the convergence of the projected results depends on the size of the initial Krylov space $M$. Fig.~\ref{fig:convergeK} shows the DMFT spectral functions with $(L,p)=(4,2)$ for $U/D=2.0$ calculated with $M$ ranging from 50 up to 400. The details of the quasiparticle peak and the upper Hubbard band are shown in the insets. For small $M$ values, small oscillations are seen on the side of the quasiparticle peak, whose amplitude decreases with increasing $M$. The line shape becomes smooth and converges between $M=300$ and 400. The size of the side peak on the Hubbard band is also seen to be $M$ dependent, which becomes static for $M \geq 100$. For all spectra shown in Fig.~\ref{fig:sf}, their convergence in $M$ is tested, which typically requires a value no more than a few hundred.

Finally, we comment on the computation cost of the proposed projection method. The most time-consuming part of the method is the generation of the initial Krylov space $\mathcal{K}^M$ and evaluating the Hamiltonian and overlap matrix element of the Krylov states as described in Sec.~\ref{sec:proj:gf}. The computation time then strongly depends on $L$ and the size of the Krylov space $M$. For $(L,p)=(1,2)$, calculating one $G(\omega)$ takes less than two minutes using a single CPU core (with $G^{\pm}(\omega)$ less than one minute each). The computation cost increases substantially with increasing $L$ due to the increase of system size, and consequently the necessary increase of $M$. For the most challenging case of $U/D=2.0$ and $(L,p)=(4,2)$, calculating one $G(\omega)$ with $M=300$ takes about two hours on a node with two eight-core processors (Intel Xeon E5-2630 v3, 2.40 GHz). However, as shown in Fig.~\ref{fig:convergeK}, the spectral function calculated with $M=100$ already closely resembles the converged result and correctly reproduces all the key features. It takes about twenty minutes to compute.

\section{Conclusion}

In conclusion, we have proposed a projection scheme for efficiently solving impurity models represented on a natural-orbital basis set. We have shown that for a one-band Hubbard model solved within DMFT, accurate Green's functions can be calculated directly on the real-frequency axis for all interaction strengths in the matter of minutes while including a few hundred bath sites. We reiterate here that although the particle-hole symmetric Bethe lattice is discussed above as a proof of concept, given the generality of the construction of natural orbitals, the proposed method applies to general fermionic impurity Hamiltonians regardless of their details. In addition, other than the DMRG plus Lanczos framework as we presented here, we expect the projection approach to work equally well with wave-function based techniques when calculating spectral functions, e.g. correction-vector method~\cite{Till1999}, dynamical DMRG~\cite{Jeckelmann2002}, and various time-evolution methods~\cite{Schollwock2005}. For multi-band problems, the method should further benefit from loop-free higher-connectivity tensor product states such as tree~\cite{Shi2006,Murg2010}, fork~\cite{Holzner2010,Bauernfeind2017}, or comb~\cite{Chepiga2019} tensor networks. As an outlook, we comment that our method can be straightforwardly extended to calculating various core-level spectroscopy starting from the converged DMFT ground state~\cite{Haverkort2014}, which can complement the conventional multiplet ligand-field calculations~\cite{Haverkort2012,Haverkort2014} that commonly have difficulties capturing effects such as resonances, edge singularities, and band excitations due to the limited degrees of freedom included in the Hamiltonian.

\section{Acknowledgment}
This work is supported by Deutsche Forschungsgemeinschaft (DFG) under Germany’s Excellence Strategy EXC-2181/1 - 390900948 (the Heidelberg STRUCTURES Excellence Cluster).

\appendix

\section{Convergence of Green's Functions in $L$ and Comparison to Exact Results}

\begin{figure}[htb]
  \includegraphics{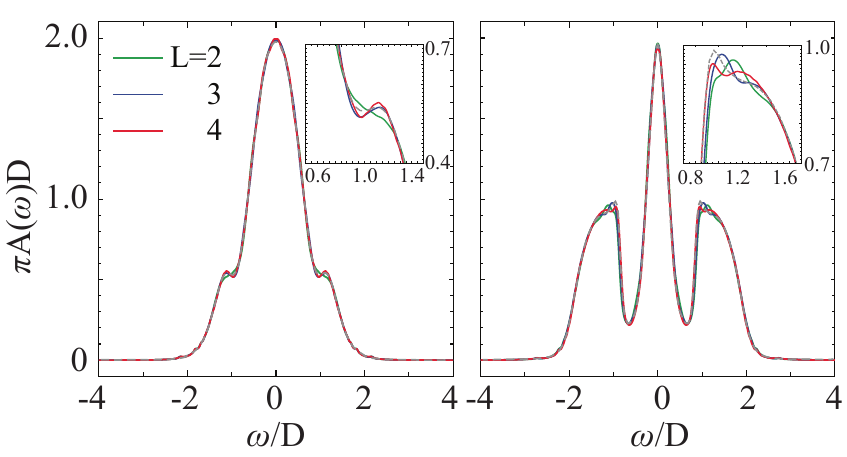}
  \caption{\label{fig:convergeL} DMFT spectral functions for (a) $U/D=1.0$ and (b) $U/D=2.0$ for $L=2$, 3, and 4 (solid lines), in comparison to exact results obtained by TEBD (dashed lines) reproduced from Fig. 1 in Ref.~\cite{Ganahl2015}. Note that the TEBD results are calculated with 119 bath sites.}
\end{figure}

Figure~\ref{fig:convergeL} shows the DMFT spectral functions obtained for $U/D=1.0$ and 2.0 with $p=2$ and $L=2,3,4$. Compared to the exact results by solving the full impurity model~\cite{Ganahl2015}, the key spectral features including the width of the quasiparticle peak and the size and position of the Hubbard bands are well reproduced already with $L=2$.

\FloatBarrier
\bibliographystyle{apsrev4-1}
\bibliography{dmrg_no.bib}

\end{document}